# Focal cortical dysplasia as a cause of epilepsy: the current evidence of associated genes and future therapeutic treatments.

Garrett L. Garner BS[1], Daniel R. Streetman BS[1], Joshua G. Fricker BS[1], Neal A. Patel BS[1], Nolan J. Brown BS[2], Shane Shahrestani MS, PhD[4,5], Julian L. Gendreau MD[3,*]

[1]*School of Medicine, Mercer University, Savannah, GA, USA*
[2]*Department of Neurosurgery, University of California Irvine, Irvine, CA, USA*
[3]*Department of Biomedical Engineering, Johns Hopkins University, Baltimore, MD, USA*
[4]*Keck School of Medicine, University of Southern California, Los Angeles, CA, USA*
[5]*Department of Medical Engineering, California Institute of Technology, Pasadena, CA, USA*
*Correspondence: jgendre1@jhu.edu (Julian L. Gendreau)

## Author Contributions

Garrett Garner retrieved data and produced the initial manuscript draft. Daniel Streetman and Joshua Fricker retrieved data. Neal Patel, Nolan Brown, Shane Shahrestani, and Julian Gendreau provided critical revision for the initial draft. All authors contributed to editorial changes in the manuscript. All authors read and approved the final manuscript.

## Ethics Approval and Consent to Participate

Not applicable.

## Acknowledgment

Not applicable.

## Funding

This research received no external funding.

## Conflict of Interest

The authors declare no conflict of interest.

**Abstract**

Focal cortical dysplasias (FCDs) are the most common cause of treatment resistant epilepsy affecting the pediatric population. Most individuals with FCD have seizure onset during the first five years of life and the majority will have seizures by the age of sixteen. Many cases of FCD are postulated to be the result of abnormal brain development in utero by germline or somatic gene mutations regulating neuronal growth and migration during corticogenesis. Other cases of FCD are thought to be related to infections during brain development, or even other causes still unable to be fully determined. Typical anti-seizure medications are oftentimes ineffective in FCD as well as surgery is unable to be successfully performed due to the involvement of eloquent areas of the brain or insufficient resection of the epileptogenic focus, posing a challenge for physicians. The genetic nature of FCD provides an avenue for drug development with several genetic and molecular targets undergoing study over the last two decades.

## 1. Introduction

Focal cortical dysplasias (FCDs) are malformations of the cerebral cortex leading to high risk of seizures and other neurological dysfunctions. This pattern of lesion was originally described in the early 1970's by Taylor et al., who attributed its defining characteristics utilizing microscopic histopathology.[1] Cortical disorganization, bizarre, large neurons and balloon neurons seen under magnification were grounds for diagnosing seizures secondary to "FCD".[2] Further nuances in FCD classification were established in 2011 with the use of a three-tiered, sub-classified system that quickly communicates extent of lesion, histopathology and an accompanying, principal lesion.[3] These distinctions, which are now readily presented in pathological reports, allow for targeted research into the various molecular underpinnings unique to each FCD subtype, as well as the novel therapeutics that exploit them. Briefly, FCD type I is characterized with malformed cortical layering, either radial (Type Ia), tangential (Type Ib) or mixed (Type Ic) layer abnormalities and is more likely to be associated with intellectual and autism spectrum disorders more than FCD type II.[4] FCD type II consists of abnormal cortical lamination, specifically, with cytologic abnormalities, either without balloon cells (Type IIa) or with balloon cells (Tape IIb). [3] In contrast to the vaster and more diffuse lesions of FCD Type I, Type II lesions are more likely to be localized in a single lobe with well-defined margins - a distinction that grants greater seizure-free outcomes post-resection.[5] While type I and type II lesions are isolated, type III lesions are defined by their association with another principal lesion, including hippocampal sclerosis (Type IIIa), glial or glioneuronal tumor (Type IIIb), adjacent vascular malformation (Type IIIc) or adjacent to lesions acquired in early life (Type IIId). A summary of the three-tiered International League Against Epilepsy (ILAE) classification system is provided in Table 1.

These malformations have been found to involve germline or somatic gene mutations that regulate neuronal growth and migration during corticogenesis of the developing embryo [6]. The exact incidence and prevalence of FCD in the general population is unknown because a definitive diagnosis of FCD requires tissue from surgical resection. However, it has been reported that up to 23% of patients receiving surgical resection of seizure foci have a FCD diagnosis [7].

Table 1: Three-tiered International League Against Epilepsy (ILAE) [3] FCD classification system

| FCD Type | Description |
| --- | --- |
| FCD Type Ia | Cortical dysplasia with abnormal radial cortical migration (i.e. vertical distance between cortical layers in perpendicular cross-section) |
| FCD Type Ib | Cortical dysplasia with abnormal tangential composition of cortical layers (i.e. poorly defined 6-layered architecture) |
| FCD Type Ic | Cortical dysplasia with abnormal radial migration and tangential cortical lamination |
| FCD Type IIa | Dysmorphic neurons without balloon cells* |
| FCD Type IIb | Dysmorphic neurons with balloon cells |
| FCD Type IIIa | Temporal cortical dyslamination with hippocampal sclerosis |
| FCD Type IIIb | Cortical dyslamination adjacent to a glial or glioneuronal tumor |
| FCD Type IIIc | Cortical dyslamination adjacent to vascular malformation |
| FCD Type IIId | Cortical dyslamination adjacent to previously acquired lesions (e.g. trauma, ischemic injury, encephalitis) |

*Balloon cells have enlarged soma with eosinophilic cytoplasm on hematoxylin and eosin (H&E) stain, lack Nissl substance, present with multiple nuclei and can be cortical or subcortical

Because of their association with genetically derived proteins and signaling pathways, FCDs typically present in the first five years of life with focal-onset seizures. Sometimes before the age of sixteen years, progression to generalized tonic-clonic seizures in these patients can occur [8, 9]. Intrinsic cortical excitability has repeatedly been demonstrated in FCDs due to altered expression of neurotransmitter receptors [6, 10]. The increased expression of many cytoskeletal elements have also been found, such as nestin, peripherin, vimentin and α-internexin [11]. The expression of doublecortin-like (DCL) is increased which plays a critical role in division, radial migration and proliferation of neuronal progenitor cells [12].

Treatment in patients with symptomatic FCDs include cation channel (Na or Ca) and/or $GABA_A$/Cl-channel modulating drugs, however very few patients with epilepsy from FCDs obtain sufficient seizure control from medications alone [6]. Therefore, treatment of FCDs represent significant clinical challenges due to their medication resistant nature. If two or more antiepileptic drugs (alone or together) fail to deliver proper seizure management, surgical resection of the FCD may be considered. However, significant seizure reduction or remission is approximately 60% after surgery [13]. FCDs in eloquent brain areas such as the posterior frontal lobe involved with motor function and the dominant inferior frontal gyrus and posterior third of the superior temporal gyrus that processes language pose significant surgical challenges and limits patient selection for surgery. Additional therapies have been hypothesized to treat seizures including the use of a ketogenic diet [14] and neuromodulators such as vagus nerve stimulators [15].

A recent consideration for treatment options is the development of targeted molecular therapies. Genetics appears to play a clear role in the pathogenesis of some FCDs, with some somatic mutations being proposed as a cause for the disease [6]. Surgical resection of FCDs have shown somatic mutations, and there are evidence from mouse models suggesting in-utero human papillomavirus (HPV) 16 may play a role in the development of somatic genetic abnormalities seen in FCDIIB [16]. Germline mutations are also associated with FCDs in families with multiple affected members [17]. Common inheritance patterns have not been reported in these families, and previous research has postulated a "two-hit" mechanism providing explanation for the lack of occurrence in healthy family members [18]. Mutations in *TSC1*, *TSC2*, and *DEPDC5* regulatory genes have been identified in familial cases of FCDs and mTOR-mediated corticogenesis is their common proliferative pathway [19]. For patients with seizures refractory to current treatment options, genetic therapy could potentially be a future option for achieving seizure relief for patients with FCDs [20].

## 2. Characterized Mutations

### 2.1 Focal Cortical Dysplasia Type 1

Several molecular markers have been found in studies and cases with patients diagnosed with FCD I on pathological specimens, however these markers for FCD I are not as robustly characterized as in FCD II (Table 2) [21-25].

**Table 2: Gene mutations for focal cortical dysplasia type I.**

| Gene | Description |
|---|---|
| *mTOR* [21] | Inconclusive on role in FCD I |
| *DEPDC5* [17, 18, 26-32] | Subunit of GATOR complex that directly inhibits mTOR activity |
| *AKT3* [33] | Activates mTOR resulting in cell growth and proliferation |
| *KCNT1* [34] | Potassium-sodium activated channel that is involved in synaptic transmission |
| *NPRL2* [22] | Inhibits mTOR activity by activating RagGTPases |
| *NPRL3* [22] | Inhibits mTOR activity by activating RagGTPases |
| *PCDH19* [35] | Neuronal migration and formation of synaptic connections |
| *SCN1A* [36] | Voltage gated sodium ion channel 1- α subunit for sodium channels |
| *SLC35A2* [37] | UDP-galactose transporter that allows galactose transport for glycosylation |
| *STXBP1* [38] | Binding protein that regulates syntaxin controlling neurotransmitter release |

*Abbreviations:* mTOR, Mammalian target of rapamycin; DEPDC5, DEP Domain Containing 5, AKT3, AKT Serine/Threonine Kinase 3; KCNT1, Potassium channel subfamily T, member 1; NPRL2, Nitrogen permease regulator 2-like; NPRL3, Nitrogen permease regulator 3-like; PCDH19, Protocadherin 19; SCN1A, sodium voltage-gated channel alpha subunit 1; SLC35A2, Solute Carrier Family 35 Member A2; STXBP1, syntaxin-binding protein 1

*2.2 Focal Cortical Dysplasia Type II*

The identified markers for FCD II are more robust, with stronger correlations that could be made for the patterns of genetic mutations found (Table 3). Patients with FCD II were most frequently studied to have somatic mTOR gene mutations[21, 22, 24-28, 30, 39-41]. *DEPDC5* is a gene which provides instructions for making a protein that is part from the protein complex GATOR1. This complex acts to inhibit the mTOR pathway.[42] *DEPDC5* has been found to be the second most studied genetic mutation, and leads to mTOR hyperactivation.

**Table 3: Gene mutations for focal cortical dysplasia type II.**

| Gene | Description |
| --- | --- |
| *mTOR* [21, 22, 24-26] | Central controller of cell growth and proliferation |
| *DEPDC5* [17, 18, 26-32] | Subunit of GATOR complex that directly inhibits mTOR activity |
| *NPRL2* [17, 18, 28-32] | Subunit of GATOR complex that inhibits mTOR activity by deactivating RagGTPases |
| *NPRL3* [17, 18, 28-32] | Subunit of GATOR complex that inhibits mTOR activity by deactivating RagGTPases |
| *TSC1* [25, 28, 43] | When activated, dephosophorylates Rheb, which inhibits mTOR |
| *TSC2* [25, 28, 43] | When activated, dephosophorylates Rheb, which inhibits mTOR |
| *AKT1* [31, 39] | Inhibits TSC1 and TSC2, Activates mTOR resulting in cell growth and proliferation |
| *AKT3* [28, 31, 44] | Inhibits TSC1 and TSC2, Activates mTOR resulting in cell growth and proliferation |
| *PTEN* [45] | Epistatically controls the kinase Akt, which controls the half-life of the hamartin-tuberin complex |
| *PIK3C2B* [30, 46] | Protein kinase involved in cell growth, signaling, and oncogenic transformation |
| *PIK3C3* [30] | Protein kinase involved in cell growth, signaling, and oncogenic transformation |
| *PIK3CA* [30] | When activated, causes aberrant mTOR activation in the excitatory neurons and glia |

*Abbreviations*: mTOR, Mammalian target of rapamycin; DEPDC5, DEP Domain Containing 5; NPRL2, Nitrogen permease regulator 2-like; NPRL3, Nitrogen permease regulator 3-like; TSC1, TSC Complex Subunit 1; TSC2, TSC Complex Subunit 1; AKT1, AKT Serine/Threonine Kinase 1; AKT3, AKT Serine/Threonine Kinase 3; PTEN, Phosphatase and tensin homolog; PIK3C2B, Phosphatidylinositol-4-Phosphate 3-Kinase Catalytic Subunit Type 2 Beta; PIK3C3, Phosphatidylinositol 3-Kinase Catalytic Subunit Type 3; PIK3CA, Phosphatidylinositol-4,5-Bisphosphate 3-Kinase Catalytic Subunit Alpha

*2.3 Focal Cortical Dysplasia Type III*

Information regarding FCD III is limited with only minimal case reports published in the literature. It is characterized by dyslamination with normal neurons similar to FCD I, however adjacent to other pathological lesions which differentiates it from FCD I [23]. Therefore, due to its rarity, molecular markers and targets for treatment are not yet well elucidated [23].

## 3. Potential Therapeutic Targets

With these identified molecular markers of FCD, several postulated strategies exist to identify targets that contribute to the formation of cortical malformations in patients with FCD. Currently, only one treatment pathway (mTOR) is undergoing clinical trials (Table 4).

**Table 4: Potential molecular therapies for the treatment of focal cortical dysplasia.**

| Molecular Therapy/Target | Study Characteristics |
|---|---|
| mTOR Inhibitors | Shown to have clinical benefit in tuberous sclerosis patients. One clinical trial is already completed finding reduced mTOR expression in patients treated with everolimus before craniotomy for seizure focus resection when compared to control (NCT02451696). Another clinical directly evaluated seizure activity with FCD after treating with everolimus. However, the results still remain unpublished (NCT03198949). |
| *HCN4* | Preclinical studies using mouse models have found reduced seizures and reduced HCN4 expression in mouse models treated with rapamycin [29]. |
| Simufilam (PTI-125) | PTI-125 treated mice have shown reduced seizure frequency by > 60% compared to vehicle-treated mice with saline [41]. |
| SV2A | Patients with treatment-resistant FCD IIB had decreased expression of SV2A on immunohistological staining. In animal models, mice with knock-out of the SV2A gene were found to have intractable seizures [47]. |
| DNMT Inhibitors | The amount of two different types of DNMT proteins were significantly elevated in human, FCD patients with temporal lobe epilepsy [48]. |

*Abbreviations: mTOR,* Mammalian target of rapamycin; HCN4, Hyperpolarization Activated Cyclic Nucleotide Gated Potassium Channel 4; SV2A, Synaptic Vesicle Glycoprotein 2A, DNMT, DNA methyltransferase

## 3.1 Mamallian target of rapamycin (mTOR)

The mTOR pathway plays a key role in cell growth and proliferation including the development of the cerebral cortex. It is hypothesized that the mTOR signaling pathway helps produce the abnormally large cells in the cortex of those with FCD as well as the abnormal the neurogenesis which will lead to epilepsy. One study demonstrated that downregulating mTOR signaling with rapamycin (an mTOR inhibitor) treatment normalized glutamatergic but not GABAergic synapses [49]. An animal model of FCD showed seizure severity and duration was strongly suppressed via mTOR pathway inhibition from rapamycin [50]. It is important to consider that withdrawal of treatment may cause recurrence of symptoms, therefore efficacy of treatment appears dependent on continuous treatment with the medication [20, 51].

As rapamycin was commonly used in animal models, current clinical trials focus on the use of everolimus, a synthetic analogue of rapamycin. Everolimus has been used to treat other cortical malformations, such as those seen in tuberous sclerosis and has shown to markedly reduce seizure frequency in these patients [52]. Only two clinical trials with medical interventions are ongoing today with FCD patients.

One trial (NCT02451696) was a phase II trial sponsored by New York University Langone Health, which has been completed and the data has been evaluated. The trial consisted of five patients undergoing Everolimus treatment for both FCD and tuberous sclerosis for 28 days prior to resection of the epileptic foci using surgery. One patient disenrolled from the trial before completion. Ten patients were used as controls and underwent surgery for the evacuation of a hematoma after suffering from hemorrhagic stroke. The primary outcome was to evaluate adverse effects of the drug, and no patients in either the treatment group or control group suffered adverse events. In addition, the mTOR protein expression was reduced when compared to controls. Outcomes of seizure frequency were not measured.

Another trial is a phase II prospective, randomized, double-blind, placebo-controlled clinical trial (NCT03198949) which evaluated the use of everolimus in patients with FCD II that continue to have seizures after using two antiepileptic drugs and surgical resection. The primary outcome was to achieve a 50% reduction rate in seizure activity by the final 4 weeks of the 65 week study. The total enrollment was 23 patients and was sponsored by Yonsei University in South Korea. It seems to be now complete and results will soon become available. With the extensive genetic component of FCDs, further clinical trials of FDA approved mTOR inhibitors such as temsirolimus, rapamycin, and sirolimus are necessary to ensure physicians have the best tools to combat medication resistant seizures in FCDs [53, 54].

In recognition of the established role mTOR/GATOR play in the FCD pathogenesis, directly modulating protein activity via gene therapy has also expectedly been a focus of research. Studies that selectively downregulate components of mTORC2 through mice with gene knock-out or antisense oligonucleotide blockade of mRNA products show decreased neuronal hypertrophy [55] and decreased seizure activity [56]. Incorporation of gene therapy into novel therapeutics is currently in early stages and limited to animal models [55, 56].

*3.2 Hyperpolarization-activated cyclic nucleotide-gated potassium channel isoform 4 (HCN4)*

HCN4 is a gated potassium channel which is not present in normal pyramidal neurons yet has been found to be abnormally expressed in rat models and human patient diseased neurons with FCD II. It is believed that ectopic HCN4 expression is a potential mechanism responsible for seizure generation in mTOR dependent focal cortical malformation. It was found to be dependent on mTOR, and blocking HCN4 in vivo with inactivated HCN4 subunits reduced seizures in rat models [57]. There are currently no HCN blockers that are selective for HCN4, however rapamycin can eliminate ectopic HCN4 expression at high doses, which can cause severe adverse effects. Targeting HCN4 specifically with potentially short hairpin RNA (shRNA)-based gene therapy for FCD II is a potential future therapy option that would eliminate the adverse effects associated with high dose rapamycin [57].

*3.3 Simufilan (PTI-125)*

A potential future therapy for FCD II is the small molecule PTI-125 to target actin cross-linking protein filamin A (FLNA) [41]. This protein serves as a scaffold for over 90 binding partners, including transcription factors, intracellular signaling molecules, channels and receptors [58]. These proteins are found to be increased in the resected cortical tissue responsible for causing seizures in patients with FCD II [41, 58]. In a Rheb$^{CA}$ mouse model, FLNA knockdown decreased seizure frequency independent of mTOR signaling [41]. The small molecule PTI-125 is a modulator of aberrant FLNA function and also currently undergoing study for Alzheimer's disease. PTI-125 treated mice have shown reduced seizure frequency by > 60% without affecting mTOR activity compared to vehicle-treated mice with saline [41]. This treatment option would provide an mTOR-independent therapeutic avenue and would likely be applicable for patients with mutations in PI3K pathway genes.

*3.4 Synaptic vesicle glycoprotein 2A (SV2A)*

SV2A is a membrane protein that is expressed in synaptic vesicles and modulates neurotransmitter release by altering synaptic action potentials. SV2A is the binding site of the levetiracetam and its analogs, and it was found that patients with treatment-resistant FCD IIB had decreased expression of SV2A on immunohistological staining [47, 59-61]. Further, SV2A-knockout mice are shown to have severe convulsive seizures [47]. Methods of enhancing SV2A function could potentially be explored [47].

3.5 *DNA methyltransferases (DNMTs) inhibitors*

DNMTs are responsible for DNA methylation of cytosine residues to 5-methylcytosine. Changes in DNMT enzyme activity and expression have been noted in patients with epilepsy [48]. It is thought that epigenetic mechanisms may be implicated in the pathogenesis of FCD, more specifically DNA methylation. DNMT proteins were found to be significantly elevated in patients with temporal lobe epilepsy in human brain tissue [48]. DNMT inhibitors could be a potential treatment for FCD by dampening the changes in the epigenomic gene expression [48, 62]. Practical problems with this approach include ubiquity of epigenetically modifiable sites and toxicities and teratogenicities associated with systematic administration of the epigenetic drug [62]. There is a need for selectivity for not only the central nervous system, but also specific regions within the central nervous system for any potential use of a DNMT inhibitor for FCD [62].

## 5. Conclusions

Our understanding of the genetic mutations and molecular targets for patients with FCD has progressed significantly in the past two decades due to advancements in laboratory technology, and resulting from the expansion of knowledge with respect to the molecular mechanisms of epilepsy. Several options exist as potential molecular therapeutic solutions for targeted therapy in these patients, with inhibition of the mTOR pathway being the currently most studied pathway in clinical trials. One previous clinical trial resulted with no significant adverse events with the mTOR inhibitors, and a second clinical trial evaluated the use of mTOR inhibitors for its potential in reducing seizures directly. However, we are still waiting for the published results.

Future aims of further clinical trials on mTOR inhibitors should include establishing an optimal dosing regimen, establishing optimal year of age for treatment initiation and establishing an ideal length of treatment. There also exists other potential strategies, including targeting cytoskeletal proteins and channel proteins that are thought to play a significant role in epileptogenesis that we hope some day will also be better approached.

**Abbreviations**

DNMT, DNA methyltransferases; DCL, doublecortin-like; FCD, Focal cortical dysplasia; HCN4, Hyperpolarization-activated cyclic nucleotide-gated potassium channel isoform 4; mTOR, Mammallian target of rapamycin; shRNA, short hairpin RNA; FLNA, actin cross-linking protein filamin A; SV2A, Synaptic vesicle glycoprotein 2A; HPV, human papillomavirus